\documentstyle[aps,prl,twocolumn,epsfig,floats]{revtex}

\newcommand{\spur}{\text{tr}\,}
\newcommand{\imagteil}{\text{Im}\,}

\begin{document}

\draft

\date{May 1999}
\title{Quantum-limited linewidth of a chaotic laser cavity}
\author{M. Patra, H. Schomerus, and C. W. J. Beenakker}
\address{Instituut-Lorentz, Universiteit Leiden, P.O. Box 9506, 2300 RA
Leiden, The Netherlands}

\twocolumn[
\widetext
\begin{@twocolumnfalse}

\maketitle

\begin{abstract}
A random-matrix theory is presented for the linewidth of a laser cavity in which
the radiation is scattered chaotically. The linewidth is enhanced above the
Schawlow-Townes value by the Petermann factor $K$, due to the
non-orthogonality of the cavity modes. The factor $K$ is expressed in terms of a
non-Hermitian random matrix and its distribution
is calculated exactly for the case that the
cavity is coupled to the outside via a small opening. The average of $K$ is
found to depend non-analytically on the area of the opening, 
and to greatly exceed the most probable value.
\end{abstract}


\vspace{0.5cm}

\narrowtext

\end{@twocolumnfalse}
]

It has been known since the conception of the
laser~\cite{schawlow:58a} that vacuum fluctuations of the
electromagnetic field ultimately limit the narrowing of the emission
spectrum by laser action. This quantum-limited linewidth, or
Schawlow-Townes linewidth,
\begin{equation}
	\delta\omega = \case{1}{2} \Gamma^2 / I\;,
	\label{startgl}
\end{equation}
is proportional to the square of the decay rate $\Gamma$ of the lasing
cavity mode~\cite{footnote1}
and inversely proportional to the output power $I$ (in
units of photons/s). Many years later it was
realised~\cite{petermann:79a,siegman:89} that the
fundamental limit is larger than Eq.~(\ref{startgl}) by a factor $K$
that characterises the non-orthogonality of the cavity modes. This
excess noise factor, or Petermann factor, has generated an extensive
literature (see the recent
papers~\cite{cheng:96a,eijkelenborg:96a,brunel:97a,grangier:98a,siegman:98a}
and references therein), both because of its fundamental significance
and because of its practical importance.

Theories of the enhanced linewidth usually factorise $K=K_l K_r$ into a
longitudinal and transverse factor, assuming that the cavity mode is
separable into longitudinal and transverse modes. Since a
longitudinal or transverse mode is essentially one-dimensional, that
is a major simplification. Separability breaks down if the cavity has
an irregular shape or contains randomly placed scatterers. In the
language of dynamical systems, one crosses over from integrable to
chaotic dynamics~\cite{haake:91}. Chaotic laser cavities have
attracted much interest recently~\cite{noeckel:97a},
but not in
connection with the quantum-limited linewidth.

In this paper we present a general theory for the Petermann factor in
a system with chaotic dynamics, and apply it to the simplest case of a
chaotic cavity radiating through a small opening. Chaotic systems
require a statistical treatment, so we compute the probability
distribution of $K$ in an ensemble of cavities with small variations
in shape and size. We find that the average of $K-1$ depends {\em
non-analytically} $\propto T \ln T^{-1}$ on the transmission probability $T$
through the opening, so that it is beyond the reach of simple perturbation
theory. The most probable value of $K-1$ is $\propto T$, hence it is
parametrically smaller than the average.

The spectral statistics of chaotic systems is described by
random-matrix theory~\cite{haake:91,mehta:90}. We begin by
reformulating the existing theories for the Petermann
factor~\cite{grangier:98a,siegman:98a} in the framework of
random-matrix theory. Modes of a closed cavity, in the absence of
absorption or amplification, are eigenvalues of a Hermitian operator
$H_0$. For a chaotic cavity, $H_0$ can be modelled by an $M\times M$
Hermitian matrix with independent Gaussian distributed elements. (The
limit $M\to\infty$ at fixed spacing $\Delta$ of the modes is taken at
the end of the calculation.) The matrix elements are real because of
time-reversal symmetry. 
(This is the Gaussian orthogonal ensemble~\cite{mehta:90}.)
A small opening in the cavity is described by
a real, non-random $M\times N$ coupling matrix $W$, with $N$ the
number of wave channels transmitted through the opening. (For an
opening of area ${\cal A}$, $N\simeq 2\pi {\cal A}/\lambda^2$ at wavelength
$\lambda$.) Modes of the open
cavity are complex eigenvalues (with negative imaginary part) of the non-Hermitian
matrix $H=H_0 - i\pi W W^T$. The scattering matrix $S$ at frequency
$\omega$ is related to $H$ by~\cite{verbaarschot:85a}
\begin{equation}
	S = \openone - 2 \pi i W^T ( \omega - H )^{-1} W \;.
	\label{Smatrix}
\end{equation}
It is a unitary and symmetric, random $N\times N$ matrix, with poles at the
eigenvalues of $H$.

We now assume that the cavity is filled with a homogeneous amplifying
medium (amplification rate $1/\tau_a$). This adds a term $i/2\tau_a$
to the eigenvalues, shifting them upwards towards the real axis. The
lasing mode is the eigenvalue $\Omega-i\Gamma/2$ 
closest to the real
axis, and the laser threshold is reached when the decay rate $\Gamma$
of this mode equals the amplification rate $1/\tau_a$~\cite{misirpashaev:98a}.
Near the laser
threshold we need to retain only the contribution from the lasing mode
(say mode number $l$) to the scattering matrix~(\ref{Smatrix}),
\begin{eqnarray}
	S_{nm} &=& - 2 \pi i (W^T U)_{nl} (\omega-\Omega
		+ i \Gamma/2-i/2\tau_a )^{-1} \nonumber\\
	& &	\qquad \mbox{} \cdot (U^{-1} W)_{lm} \;,
	\label{Smatrix2}
\end{eqnarray}
where $U$ is the matrix of eigenvectors of $H$. Because $H$ is a real
symmetric matrix, we can choose $U$ such that $U^{-1}=U^T$ and write
Eq.~(\ref{Smatrix2}) in the form
\begin{equation}
	S_{nm} = \sigma_n \sigma_m (\omega -\Omega
	+ i \Gamma/2 -i/2\tau_a)^{-1} \;,
\end{equation}
where $\sigma_n = (-2\pi i)^{1/2} (W^T U)_{nl}$ is the complex
coupling constant of the lasing mode $l$ to the $n$-th wave channel.

The Petermann factor $K$ is given by
\begin{equation}
	\sqrt{K} = \frac{1}{\Gamma} \sum_{n=1}^N |\sigma_n|^2
	= (U^\dagger U)_{ll} \;.
	\label{Kgl1}
\end{equation}
The second equality follows from the definition of
$\sigma_n$~\cite{footnote2}, and is the matrix analogon of
Siegman's non-orthogonal mode
expression~\cite{siegman:89}. The first equality follows
from the definition of $K$ as the factor multiplying the
Schawlow-Townes linewidth~\cite{footnote3}. One verifies that $K\ge1$ because
$(U^\dagger U)_{ll}\ge (U^T U)_{ll}=1$.


The relation (\ref{Kgl1}) serves as the starting point for a calculation of the
statistics of the Petermann factor in an ensemble of chaotic cavities. Here we
restrict ourselves to the case $N=1$ of a single wave channel, 
leaving the multi-channel case for future investigation.
For $N=1$ the coupling matrix
$W$ reduces to a vector $\vec{\alpha}=(W_{11}, W_{21},\ldots,W_{M1})$.
Its magnitude $|\vec{\alpha}|^2=(M\Delta/\pi^2) w$, where $w\in [0,1]$ is related to the
transmission probability $T$ of the single wave channel by $T=4 w(1+w)^{-2}$.
We assume a basis in which $H_0$ is diagonal (eigenvalues $\omega_q$).

If the opening is much smaller than a wavelength, then a perturbation theory in
$\vec{\alpha}$ seems a natural starting point. To leading order one finds
\begin{equation}
	K = 1 + (2 \pi \alpha_l)^2 \sum_{q\ne l}
	\frac{\alpha_q^2}{(\omega_l-\omega_q)^2} \;.
	\label{Kgl2}
\end{equation}
The frequency $\Omega$ and decay rate $\Gamma$ of the lasing mode are given
by $\omega_l$ and $2\pi \alpha_l^2$, respectively, to leading order in
$\vec{\alpha}$. We seek the average $\langle K \rangle_{\Omega,\Gamma}$ of
$K$ for
a given value of $\Omega$ and $\Gamma$. The probability to find an eigenvalue at
$\omega_q$ given that there is an eigenvalue at $\omega_l$ vanishes {\em
linearly} for small $|\omega_q-\omega_l|$, as a consequence of eigenvalue
repulsion constrained by time-reversal symmetry. Since the expression
(\ref{Kgl2}) for $K$ diverges {\em quadratically} for small
$|\omega_q-\omega_l|$, we conclude that $\langle K \rangle_{\Omega,\Gamma}$ does
not exist in perturbation theory. This severely complicates the problem.

We have succeeded in obtaining a finite answer for the average Petermann factor
by starting from the exact relation
\begin{equation}
	U_{ql} z_l = \omega_q U_{ql} - i \pi \alpha_q \sum_{p} \alpha_p U_{pl}
\end{equation}
between the complex eigenvalues $z_q$ of $H$ and the real eigenvalues $\omega_q$
of $H_0$. Distinguishing between $q=l$ and $q\ne l$, and defining $d_q =
U_{ql}/U_{ll}$, we obtain two recursion relations,
\begin{mathletters}
\begin{eqnarray}
	z_l &=& \omega_l - i \pi \alpha_l^2
		- i \pi \alpha_l \sum_{q\ne l} \alpha_q d_q\;,	\\
	i d_q &=& \frac{\pi \alpha_q}{z_l - \omega_q} \biggl( \alpha_l 
		+ \sum_{p\ne l} \alpha_p d_p \biggr)\;.
	\label{tempeq13b}
\end{eqnarray}
\end{mathletters}
The Petermann factor of the lasing mode $l$ follows from
\begin{equation}
	\sqrt{K} = \biggl( 1 + \sum_{q\ne l} |d_q|^2\biggr)
		\biggl|1+\sum_{q\ne l} d_q^2 \biggr|^{-1} \;.
	\label{tempeq14}
\end{equation}

We now use the fact that $z_l$ is the eigenvalue closest to the real axis. We
may therefore assume that $z_l$ is close to the unperturbed value $\omega_l$
and replace the denominator $z_l-\omega_q$ in Eq.~(\ref{tempeq13b}) by
$\omega_l-\omega_q$. That decouples the two recursion
relations, which may then be solved
in closed form,
\begin{mathletters}
\begin{eqnarray}
	z_l &=& \omega_l - i \pi \alpha_l^2 \left( 1 + i \pi A \right)^{-1}
		\;,\\
	i d_q &=& \frac{\pi \alpha_q \alpha_l}{\omega_l-\omega_q}
		\left(1 + i \pi A \right)^{-1} \;.
\end{eqnarray}
\end{mathletters}
We have defined $A=\sum_{q\ne l} \alpha_q^2 (\omega_l-\omega_q)^{-1}$. The decay
rate of the lasing mode is 
\begin{equation}
	\Gamma = - 2 \,\imagteil z_l = 2 \pi \alpha_l^2 (1+\pi^2 A^2)^{-1}\;.
	\label{eq21}
\end{equation}
Since the lasing mode is close to the real axis, we may
linearise the expression (\ref{tempeq14}) for $K$ with respect to $\Gamma$,
\begin{equation}
	K = 1 + 4 \sum_{q\ne l} (\imagteil d_q)^2
	=  1 + \frac{(2 \pi \Gamma/\Delta) B}{1+\pi^2 A^2} \;,
	\label{tempeq23}
\end{equation}
with $B=\Delta\sum_{q\ne l} \alpha_q^2 (\omega_l - \omega_q )^{-2}$.

The conditional average of $K$ at given $\Gamma$ and $\Omega$ can be written as
the ratio of two unconditional averages,
\begin{mathletters}
\begin{eqnarray}
	\langle K \rangle_{\Omega,\Gamma} & = & 1 + (2 \pi \Gamma/\Delta)
		\langle B (1+\pi^2 A^2)^{-1} Z \rangle / \langle Z \rangle \;,\\
	Z & = & \delta(\Omega-\omega_l) \delta\left( 
		\Gamma - 2 \pi \alpha_l^2 ( 1 + \pi^2 A^2 )^{-1} \right) \;.
	\label{tempeq15}
\end{eqnarray}
\end{mathletters}
In principle one should also require that the decay rates of modes $q\ne l$ are
bigger than $\Gamma$, but this extra condition becomes irrelevant for
$\Gamma\to0$. For $M\to\infty$ the distribution of $\alpha_q$ is Gaussian
$\propto \exp(-\case{1}{2} \alpha_q^2 \pi^2 / w \Delta)$~\cite{mehta:90}.
The average of $Z$ over $\alpha_l$ yields a factor $(1+\pi^2 A^2)^{1/2}$,
\begin{equation}
	\langle K \rangle_{\Omega,\Gamma}
	 = 1 + (2 \pi \Gamma /\Delta)\frac{\langle B ( 1 + \pi^2 A^2 )^{-1/2} \rangle}{
		\langle ( 1 + \pi^2 A^2 )^{1/2} \rangle } \;,
	\label{tempeq16}
\end{equation}
where only the averages over $\alpha_q$ and $\omega_q$ ($q\ne l$) remain,
at fixed $\omega_l=\Omega$.

\begin{figure}
\epsfig{file=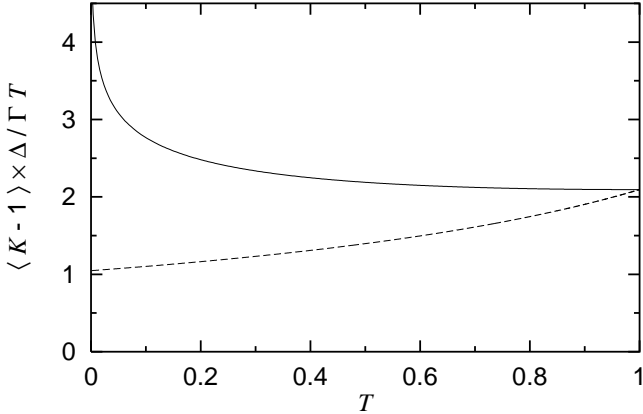,width=3.4in}
\caption{Average Petermann factor $K$ for a chaotic cavity having an
opening with transmission probability $T$. The average is performed at fixed
decay rate $\Gamma$ of the lasing mode, assumed to be much smaller than the mean
modal spacing $\Delta$. The solid curve is the result~(\ref{tempeq22})
in the presence of time-reversal symmetry, 
the dashed curve is the result~(\ref{eq24}) for broken time-reversal symmetry.
For small $T$, the solid curve diverges $\propto \ln T^{-1}$ while the dashed
curve has the finite limit of $\pi/3$. For $T=1$ both curves reach the value
$2\pi/3$.
}
\label{peter1fig}
\end{figure}

The problem is now reduced to  a calculation of
the joint probability distribution $P(A,B)$. 
This is a technical challenge, similar to the level curvature problem of
random-matrix theory~\cite{oppen:9x,fyodorov:95a}. The calculation
will be presented elsewhere, here we only give the result:
\begin{equation}
	P(A,B)=
	\frac16 \sqrt{
	\frac{\pi}{2w}}
	\frac{\pi^2A^2+w^2}{B^{7/2}}
	\exp\left[-\frac{w}{2 B}\left(\frac{\pi^2
	A^2}{w^2}+1\right)
	\right]
	\;. \label{tempeq24}
\end{equation}
Together with Eq.~(\ref{tempeq16}) this gives the mean Petermann factor
\begin{equation}
\left\langle K\right\rangle_{\Omega,\Gamma}=1-\frac{\Gamma}{\Delta}
	\frac{2\pi}{3} \frac{G^{22}_{22}\left(
		w^2\left|\begin{array}{cc}0 & 0\\ -\case{1}{2} & -\case{1}{2}
	\end{array}\right.\right)
	}{
	G^{22}_{22}\left(
	w^{2}\left|\begin{array}{cc}
	-\case 12 &\case 12\\-1 & 0 
	\end{array}\right.\right)}
	\;,
	\label{tempeq22}
\end{equation}
in terms of the ratio of two Meijer $G$-functions. 
We have plotted the result in Fig.~\ref{peter1fig}, as a function of $T=4
w(1+w)^{-2}$.

The non-analytic dependence of the average $K$ on $T$ (and hence on the area of
the opening~\cite{footnote4}) is a striking feature of our result.
For $T\ll 1$, the average reduces to
\begin{equation}
	\langle K \rangle_{\Omega,\Gamma} = 1 + \frac{\pi}{6} 
		\frac{T \Gamma}{\Delta} \ln \frac{16}{T} \;.
	\label{tempeq21}
\end{equation}

The non-analyticity results from the
relatively weak eigenvalue repulsion in the presence of time-reversal symmetry.
If time-reversal symmetry is broken by a magneto-optical effect (as in
Refs.~\cite{alt:95a,stoffregen:95a}), then the stronger quadratic repulsion is
sufficient to overcome the $\omega^{-2}$ divergence of perturbation theory and
the average $K$ becomes an analytic function of $T$. For this case, we find
instead of Eq.~(\ref{tempeq16}) the simpler expression
\begin{equation}
        \langle K \rangle_{\Omega,\Gamma}
         = 1 + ( 2 \pi \Gamma / \Delta) \frac{\langle B \rangle}{
                \langle 1 + \pi^2 A^2 \rangle } 
	\;. \label{tempeq20}
\end{equation}
Using the joint probability distribution
\begin{equation}
	P(A,B)=\frac{\left(\pi^2 A^2+w^2\right)^2}{3 w B^5}
	\exp\left[-\frac{w}{B}\left(\frac{\pi^2 A^2}{w^2}+1\right)\right]
	\;,
\end{equation}
we find the mean $K$,
\begin{equation}
	\langle K \rangle_{\Omega,\Gamma}
	= 1 + \frac{\Gamma}{\Delta} \frac{4 \pi w}{3 (1+w^2)} \;,
	\label{eq24}
\end{equation}
shown dashed in Fig.~\ref{peter1fig}. It is
equal to $\langle K \rangle_{\Omega,\Gamma}=1+\case{1}{3} \pi T \Gamma / \Delta$
for $T\ll 1$.

\begin{figure}[b!]
\epsfig{file=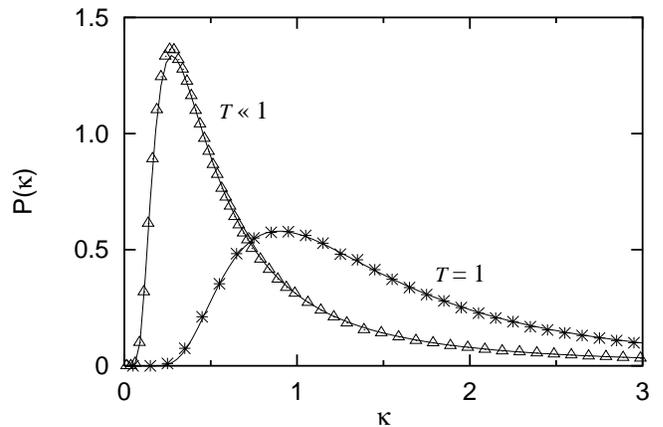,width=3.4in}
\caption{Probability distribution of the rescaled Petermann factor
$\kappa = (K - 1)\Delta / \Gamma T$ 
for $T=1$ and $T\ll 1$. The solid curves follow from
Eqs.~(\ref{distfunk}) and (\ref{tempeq19}).
The data points follow from a numerical simulation of the random-matrix model.
}
\label{figeinsklein}
\end{figure}

So far
we have concentrated on the average Petermann factor, but from 
Eqs.~(\ref{eq21}), (\ref{tempeq23}), and (\ref{tempeq24}) we can
compute the entire probability distribution of $K$ at fixed
$\Gamma$.
We define $\kappa=(K-1)\Delta/\Gamma T$.
A simple result for $P(\kappa)$ follows for $T=1$,
\begin{equation}
P(\kappa)=
	\frac{4 \pi^2}{3}\kappa^{-7/2}\exp(-\pi/\kappa) \;,
	\label{distfunk}
\end{equation}
and for $T\ll 1$,
\begin{equation}
        P(\kappa)=\frac{\pi}{12 \kappa^2}\left(1+\frac{\pi}{2\kappa}\right)
        \exp\left(-\case{1}{4}\pi / \kappa\right)
        \;, \quad \kappa T \lesssim 1 \;.
        \label{tempeq19}
\end{equation}
As shown in Fig.~\ref{figeinsklein}, both distributions are very broad and
asymmetric, with a long tail towards large $\kappa$~\cite{footnote5}. The most
probable (or modal) value of $K-1\simeq T\Gamma/\Delta$ is parametrically
smaller than the mean value (\ref{tempeq21}) for $T\ll 1$.

To check 
our analytical results 
we have also done a numerical simulation of the random-matrix model,
generating a large number of random matrices $H_0$ and computing
$K$ from Eq.~(\ref{Kgl1}). As one can see from Fig.~\ref{figeinsklein}, 
the agreement with Eqs.~(\ref{distfunk}) and (\ref{tempeq19}) is flawless.


In conclusion, we have shown that chaotic scattering causes large statistical
fluctuations in the quantum-limited linewidth of a laser cavity. We have
examined in detail the case that the coupling to the cavity is via a single
wave channel, but our random-matrix model applies more generally to coupling via
an arbitrary number $N$ of wave channels. 
We have computed exactly
the distribution of the Petermann factor for $N=1$. 
It remains an open problem to do the same for $N>1$.
This problem is related to
several recent studies of the statistics of eigenfunctions of non-Hermitian
Hamiltonians~\cite{chalker:98a,janik:99a}, but is complicated by the constraint that
the corresponding eigenvalue is the closest to the real axis.
Our study of a system with a fully chaotic phase space complements previous
theoretical work on systems with an integrable dynamics. Chaotic laser cavities
of recent experimental interest~\cite{gmachl:98a} have a
phase space that includes both integrable and chaotic regions. The study 
of the quantum-limited linewidth of such
mixed systems is a challenging problem for future research.

We have benefitted from discussions with P.\ W.\ Brouwer, K.\ M.\ Frahm,
Y.\ V.\ Fyodorov, and F.\ von\ Oppen.
This work was supported by the Dutch Science Foundation NWO/FOM and by the
TMR program of the European Union.


\begin{thebibliography}{10}

\bibitem{schawlow:58a}
A.~L. Schawlow and C.~H. Townes, Phys.~Rev. {\bf 112},  1940  (1958).

\bibitem{footnote1}
It is assumed that $\Gamma$ is much less than the linewidth
of the atomic transition and also that the lower level of the transition is
unoccupied.

\bibitem{petermann:79a}
K. Petermann, IEEE J. Quantum Electron. {\bf 15},  566  (1979).

\bibitem{siegman:89}
A.~E. Siegman, Phys.~Rev.~A {\bf 39},  1253  (1989); 
Phys.~Rev.~A {\bf 39},  1264  (1989).

\bibitem{cheng:96a}
Y.-J. Cheng, C.~G. Fanning, and A.~E. Siegman, Phys.~Rev.~Lett. {\bf 77},  627
  (1996).

\bibitem{eijkelenborg:96a}
M.~A. van Eijkelenborg, {\AA}.~M. Lindberg, M.~S. Thijssen, and J.~P. Woerdman,
  Phys.~Rev.~Lett. {\bf 77},  4314  (1996).

\bibitem{brunel:97a}
M. Brunel, G. Ropars, A. {Le Floch}, and F. Bretenaker, Phys.~Rev.~A {\bf 55},
  4563  (1997).

\bibitem{grangier:98a}
P. Grangier and J.-P. Poizat, Eur.~Phys.~J.~D {\bf 1},  97  (1998).

\bibitem{siegman:98a}
A.~E. Siegman, Phys.~Rev.~A, to be published.

\bibitem{haake:91}
F. Haake, {\em Quantum Signatures of Chaos} (Springer, Berlin, 1991).

\bibitem{noeckel:97a}
J.~U. N{\"o}ckel and A.~D. Stone, Nature {\bf 385},  45  (1997).

\bibitem{mehta:90}
M. Mehta, {\em Random Matrices} (Academic, New York, 1990).

\bibitem{verbaarschot:85a}
J.~J.~M. Verbaarschot, H.~A. Weidenm{\"u}ller, and M.~R. Zirnbauer, Phys.~Rep.
  {\bf 129},  367  (1985).

\bibitem{misirpashaev:98a}
T.~S. Misirpashaev and C.~W.~J. Beenakker, Phys.~Rev.~A {\bf 57},  2041
  (1998).

\bibitem{footnote2}
To prove the second equality in Eq.~(\ref{Kgl1}),
write $\sum_n |\sigma_n|^2 = 2 \pi (U^\dagger W W^T U)_{ll}
= 2 i (U^\dagger (H - H_0) U)_{ll}
= 2 i (U^\dagger U)_{ll} (\omega_0-i\Gamma/2)-2 i(U^\dagger H_0
U)_{ll}$ and take the real part.

\bibitem{footnote3}
For the first equality in
Eq.~(\ref{Kgl1}), write the linewidth $\delta\omega=\Gamma-1/\tau_a$
in terms of the output power $I=\int \spur S S^\dagger d \omega / 2 \pi
= ( \sum_n |\sigma_n|^2)^2 (\Gamma-1/\tau_a)^{-1} = K \Gamma^2
/\delta\omega$. The linewidth differs from the Schawlow-Townes value
(\ref{startgl}) by a factor $2 K$. The extra factor
$2$ arises from the suppression of amplitude fluctuations in
the non-linear regime above the laser threshold, as explained by
P. Goldberg, P.~W. Milonni, and B. Sundaram, Phys.~Rev.~A {\bf 44}, 1969
(1991).

\bibitem{oppen:9x}
F. {von Oppen}, Phys.~Rev.~Lett. {\bf 73},  798  (1994);
Phys.~Rev.~E {\bf 51},  2647  (1995).

\bibitem{fyodorov:95a}
Y.~V. Fyodorov and H.-J. Sommers, Z.~Phys.~B {\bf 99},  123  (1995).

\bibitem{footnote4}
The transmission probability $T$ is related to the area 
${\cal A}$ of the opening by $T\simeq{\cal A}^3/\lambda^6$ for $T\ll 1$. See H.
A. Bethe, Phys. Rev. {\bf 66}, 163 (1944).

\bibitem{alt:95a}
H. Alt, H.-D. Gr{\"a}f, H.~L. Harney, R. Hofferbert, H. Lengeler, A. Richter,
  P. Schardt, and H.~A. Weidenm{\"u}ller, Phys.~Rev.~Lett. {\bf 74},  62
  (1995).

\bibitem{stoffregen:95a}
U. Stoffregen, J. Stein, H.-J. St{\"o}ckmann, M. Ku{\'s}, and F. Haake,
  Phys.~Rev.~Lett. {\bf 74},  2666  (1995).

\bibitem{footnote5}
For the case of broken time-reversal symmetry, we find
	$P(\kappa)=\case{8}{3}\pi^4\kappa^{-5} \exp(-2\pi/\kappa)$
for $T=1$, and
	$P(\kappa)=\case{1}{6} \pi 2^{-1/2}\kappa^{-5/2}
	\left[\case{3}{4} +\pi/2\kappa+\left(\pi/2\kappa\right)^2
	\right]
	\exp\left(-\pi/2\kappa\right)$
for $T\ll 1$.

\bibitem{chalker:98a}
J.~T. Chalker and B. Mehlig, Phys.~Rev.~Lett. {\bf 81},  3367  (1998).

\bibitem{janik:99a}
R.~A. Janik, W. Noerenberg, M.~A. Nowak, G. Papp, and I. Zahed, preprint
  (cond-mat/9902314).

\bibitem{gmachl:98a}
C. Gmachl, F. Capasso, E.~E. Narimanov, J.~U. N{\"o}ckel, A.~D. Stone, J.
  Faist, D.~L. Sivco, and A.~Y. Cho, Science {\bf 280},  1556  (1998).

\end{thebibliography}

\end{document}